\documentclass{aa}
\usepackage[varg]{txfonts}
\usepackage{color}

\begin{document}

\titlerunning{Broadband nuclear emission in radio-loud BAL quasars}

\title{Broadband nuclear emission in two radio-loud BAL quasars}

\author{M. Kunert-Bajraszewska\inst{1}
     \and K. Katarzy\'nski\inst{1}
     \and A. Janiuk\inst{2}}
     
     \offprints{M. Kunert-Bajraszewska, \email{magda@astro.uni.torun.pl}}
         
\institute{Toru\'n Centre for Astronomy, Faculty of Physics, Astronomy and Informatics, NCU, Grudzi\k{a}dzka 5, 87-100 Toru\'n, Poland,
      \and Centre for Theoretical Physics, Polish Academy of Sciences, Al. Lotnik\'ow 32/46, 02-668 Warsaw, Poland}

\abstract{}{
We present modelling and interpretation of the continuum broadband emission of two broad absorption line (BAL)
quasars. The X-ray weakness of BAL quasars in comparison to non-BAL objects is possibly caused by the 
absorption of X-ray emission by the shielding material near the equatorial plane. On the other hand,
the radio-loud BAL quasars are more X-ray loud than the radio-quiet ones. This suggests that part of the X-ray emission
may arise from the radio jet. To investigate this possibility, we modelled 
the nuclear spectra of two BAL quasars in the whole available energy range.}
{We focus on the emission from the very centres of these two objects, not greater than several parsecs. The source of emission was approximated by a single, homogeneous
component that produces synchrotron and inverse-Compton radiation. The simplicity of the model
allowed us to estimate the basic physical parameters of the emitting regions, using a universal analytic approach. Such methods have already been proposed to estimate basic physical parameters in 
blazars. For the first time, in a simplified form we propose this solution for quasars.
In addition, we modelled the radiation spectra
of the accretion disk and its corona to compare them with the jets' spectra.}
{We find that in the case of radio and X-ray luminous high-redshift object 3C270.1, the nuclear X-ray continuum is dominated by the
non-thermal, inverse-Compton emission from the innermost parts of the radio jet. However, the radio core of the lobe-dominated PG\,1004+130 
is probably too weak to produce significant part of the observed X-ray emission. 
A large contribution from the X-ray emitting accretion disk and corona is produced in our
model for a sufficiently high mass of the black hole.
However, it then exceeds the observed flux. Because the large intrinsic absorption was
postulated recently by the NuSTAR observations,
we propose that the disk-corona component may still account for the X-rays produced in this
source. This part of the spectrum must nevertheless
be dominated by the X-ray jet. The results of our modelling show that the jet-linked X-ray emission is present in both strong and weak radio sources, but its fraction seems to scale with the radio jet power.
}
{}


\maketitle

\section{Introduction}

Extremely fast outflows (reaching 0.3c) of highly ionized plasma that are launched 
in the proximity of a supermassive black hole are possibly the origin of the blue-shifted 
broad  absorption lines (BAL)
in the UV part of the quasar spectrum. They have been under scrutiny for over 20 years. Together with powerful
jets, they provide a very efficient mechanism for the mass transfer from the accretion disk.
Traditionally, BALs are defined to have $\rm C_{IV}$ absorption troughs at 
least $\rm 2000\,km\,s^{-1}$ wide and are quantified by the balnicity index (BI) introduced by \citet{Weymann1991}.
When using this definition, the fraction of BAL quasars among the whole quasar population is only $\sim 15\%$ 
\citep{hewett2003, trump2006, dai08, maddox08, knigge2008, gibson2009}. 

Such a low value can be explained by
the theoretical model of \citet{elvis00} in which the outflows manifest as BALs in the quasar spectrum only if they are
seen under specific, middle-range inclination angles. In the accretion disk wind model by \citet{Murray1995}, the wind is 
launched from the disk at much lower inclination angles. Finally, the radio variability studies of some of the BAL quasars 
imply that there are polar outflows from the inner regions of a thin disk \citep{zhou2006, ghosh2007}.
Nevertheless, according to the radio studies of BAL quasars \citep{shankar2008, kun14}, the majority of them have low radio 
luminosities and moderate orientations. The observed X-ray weakness of some of them agrees with 
the above findings.

Generally, the X-ray and UV properties of BAL quasars are consistent with those of unabsorbed quasars \citep{gallagher2006, stalin2011}, but
both the X-ray and UV absorption are complex in these objects. Several other studies, however, find that after accounting
for intrinsic absorption \citep{luo13} or magnification of X-ray emission viewed from large inclination angles \citep{leah14}, 
BAL quasars are X-ray weak compared to non-BAL objects. Moreover, the degree of X-ray weakness for radio-loud and radio-quiet
BAL quasars is different, with the radio-loud ones being less X-ray weak than radio-quiet ones, compared to non-BAL quasars \citep{miller2009}.
A simple scenario suggested by a few authors says that the X-ray continuum of radio-loud BAL quasars is a superposition of disk/corona 
and small-scale jet X-ray emission \citep{wang2008, Kunert09, miller2009}.

In this work, we explore the above scenario by modelling the spectra of two radio-loud BAL quasars
in the framework of an optically thick disk plus corona model and the emission of a jet.
Both these quasars,  PG\,1004+130 and 3C270.1, are classified as large-scale
(linear size $>$20\,kpc), radio-loud AGNs, and their radio structures are well resolved with VLA,
showing central component, radio core, and jets or lobes, on
both sides. Their rather symmetric morphologies suggest a some line-of-sight  
beaming in the jets and cores, which is an important property when
interpreting the results of our modelling. In the case of PG\,1004+130, the
angle between the observer and the jet axis has been estimated to be
$\gtrsim 45^{o}$ \citep{wills99}, and the radio core is weak in this source.
Quasar 3C270.1 is a core-dominated object. The {\it Chandra} X-ray
observations also resolved the structure of both quasars. The X-ray and
radio properties of the outer, kiloparsec-scale jets, together with jet spectral-energy-distribution (SED) modelling, has been
described in detail by \citet{miller} and \citet{wilkes}.
Here we focus on the central regions of
both sources, where most of the X-ray emission is concentrated.

\begin{table}
\caption[]{Basic observational parameters}
\begin{tabular}{@{}l c c c c @{}}
\hline
\multicolumn{1}{l}{} &
\multicolumn{1}{c}{3C270.1} &
\multicolumn{1}{c}{Ref.}&
\multicolumn{1}{c}{PG\,1004+130}&
\multicolumn{1}{c}{Ref.}\\
\hline
RA (J2000)           &$12^{\rm h}20^{\rm m}33.9^{\rm s}$ & &$10^{\rm
h}07^{\rm m}26.1^{\rm s}$ & \\
Dec (J2000)          & $+33^{\rm o}43$\arcmin$12$\arcsec$$& & $+12^{\rm
o}48$\arcmin$56$\arcsec$$& \\
Redshift {\it z}     &1.53                 &   & 0.24                  & \\
P(408\,MHz)          & 280                 & 2 & $-$                   & \\
L(1.4\,GHz)          & 210                 & 2 & $15.8\pm 0.4$         & 1\\
C(5\,GHz)            & 190                 & 2 & $27.3\pm 0.2$         & 1\\
X(8.4\,GHz)          & $107.6\pm 0.1$      & 1 & $19.2\pm 0.3$         & 1\\
U(15\,GHz)           & $88.9\pm 0.4$       & 1 & $-$                   & \\ 
K(22\,GHz)           & $-$                 &   & $57.9\pm 2.0$         & 1\\
X-ray                & $6\pm 0.22$  & 3 & $4.5\pm 0.10$    & 4
\\
                     &(0.3-8\,keV)         &   &(0.5-8\,keV)           & \\  
photon index $\gamma$& 1.66                & 3 & 1.50                & 4\\ 
log$L_{\rm bol}/L_{\rm Edd}$ &0.34                & 5 &0.09                  & 6\\ 
log$M_{\rm BH}/M_{\odot}$&9.10                 & 5 &9.30                   & 6\\ 
BI                 & 52.5                & 7 &     850 & 8 \\
\hline
\end{tabular}

\vspace{0.5cm}
The radio and X-ray fluxes are related to the radio core {\it C}
(Fig.~\ref{images}) and are given in the units of mJy and $\rm
10^{-13}~erg~s^{-1}~cm^{-2}$, respectively.\\
The Balnicity index (BI) is given in the units of $\rm km\,s^{-1}$.\\
References: (1) this paper, (2) \citet{stocke}, (3) \citet{wilkes}, (4)
\citet{miller}, (5) \citet{shen}, (6) \citet{luo13}, (7) \citet{gibson2009}, 
(8) \citet{wills99}.
\label{fluxy}
\end{table}  

\section{Multi-wavelength data}

We collected the available radio, IR, optical, and X-ray data from the literature
and from the {\it Sloan Digital Sky Survey} (SDSS) database
for both PG\,1004+130 and 3C270.1 quasars. The IR and optical data of
both host galaxies can be found in the NASA/IPAC
Extragalactic Database (NED). 
The {\it Spitzer/IRAC} 3.6, 4.5, 5.8, 8, 16, and 24 $\mu m$ flux densities of 3C270.1
were taken from \citet{drouart2012}. {\it NuSTAR} X-ray observations come from the work of
\citet{luo13}.
However, the radio and {\it Chandra} X-ray fluxes in
our modelling refer only to the central part of each quasar - the
radio core. The radio and X-ray core fluxes are gathered in
Table~\ref{fluxy} for convenience. 
The 8\,GHz flux density of 3C270.1 (project AB796)
and
the 5\,GHz (project AK298), 8\,GHz (project AW249), and
22\,GHz (project AP210) flux densities of PG\,1004+130 were measured in VLA
images produced by the VLA data calibration pipeline in AIPS. The 1.4\,GHz core flux density of
PG\,1004+130 was measured in the {\it Faint Images of the Radio Sky at Twenty-cm} (FIRST) image. 
The final 8\,GHz images of both
PG\,1004+130 and 3C270.1 quasars are presented here in Fig.~\ref{images},
where the radio core is indicated as {\it C}.   

Throughout the paper, we have assumed cosmological parameters 
${\rm
H_0}$=71${\rm\,km\,s^{-1}\,Mpc^{-1}}$, $\Omega_{M}$=0.27,
$\Omega_{\Lambda}$=0.73.

\section{Notes on selected objects}

\begin{figure}
\centering
\includegraphics[width=7cm, height=7cm]{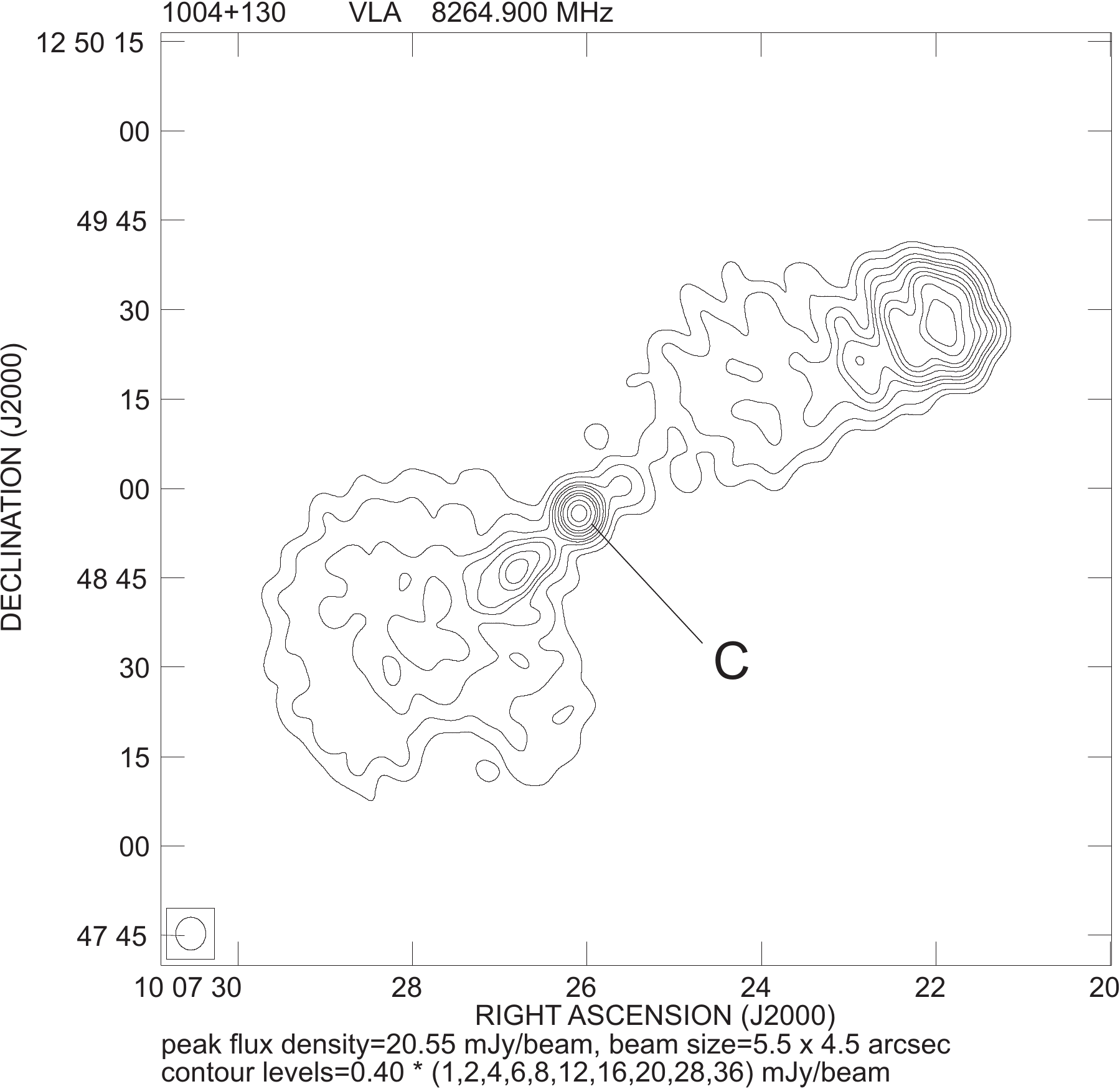}
\includegraphics[width=7cm, height=7cm]{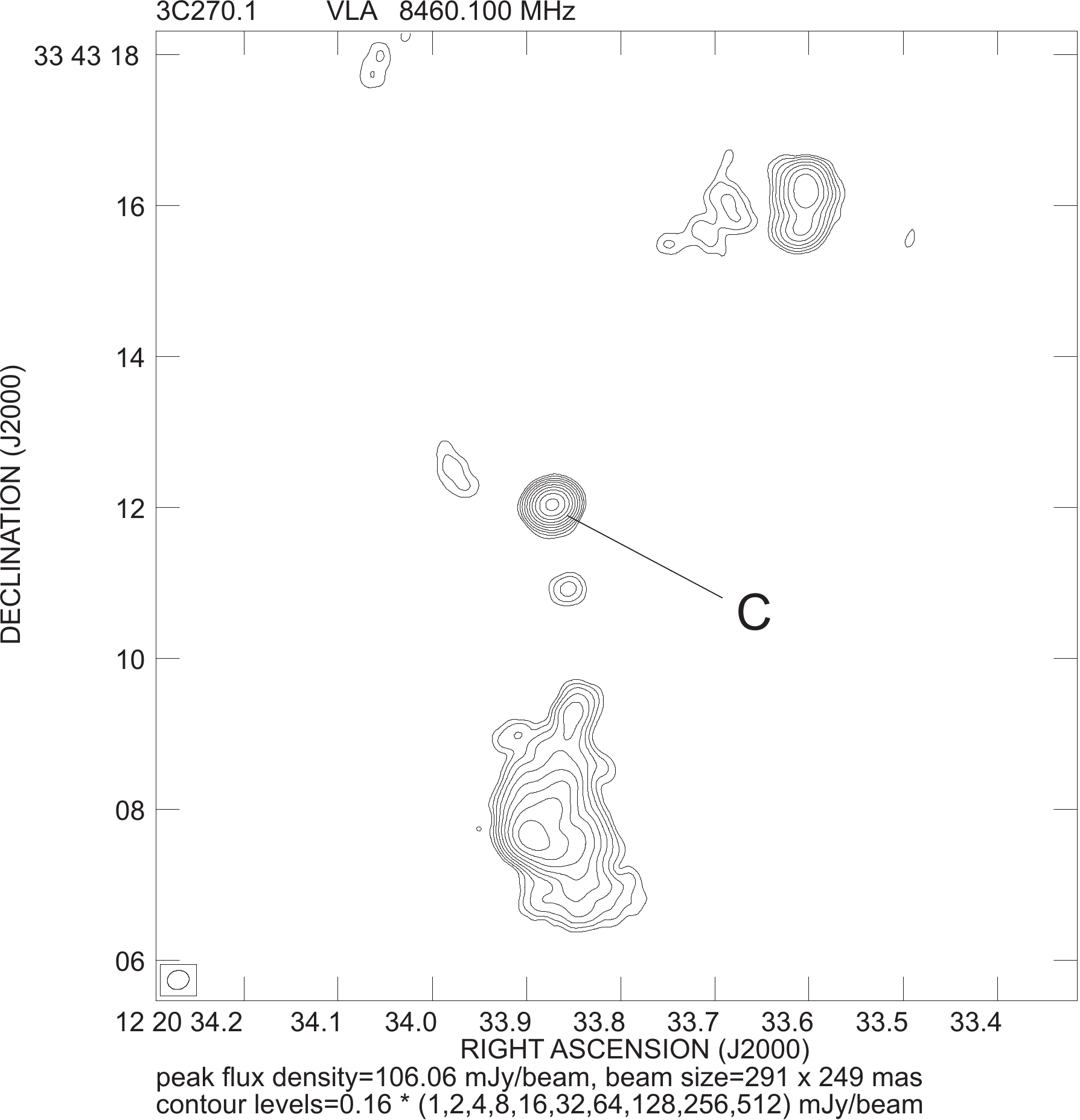} 
\caption{VLA X-band images of PG\,1004+130 (top) and 3C270.1 (bottom).
Contours increase
by a factor 2, and the first contour level corresponds to $\approx 3\sigma$.}
\label{images}
\end{figure}

{\bf PG\,1004+130.} This is a low-redshift (z=0.24), broad absorption
line
quasar \citep{wills99} with a hybrid FR\,I/FR,II radio morphology
\citep{gopwii}. The south-eastern FR\,I-like part of the source consists of
a faint lobe and a jet, which progressively fades away from the nucleus {\it C}. The
north-western part is dominated by strongly edge-brightened radio lobe,
which is typical of the FR\,II sources (Fig.~\ref{images}), which is also the brightest
component of the whole source. The overall linear size of the
source is $\sim$406\,kpc. The {\it Chandra}
observations revealed X-ray emission aligned with the FR\,I jet and X-ray
absorption variability \citep{miller}. Most of the X-ray flux is associated with the radio core.
Recently, \citet{luo13} have reported detecting this source in the softer {\it NuSTAR} band.

\noindent
{\bf 3C270.1.}
This is a high redshift (z=1.53), radio-loud broad absorption line
quasar \citep{gibson2009}. It has been classified as a 'dog-leg' or bent quasar
\citep{stocke}. The radio structure consists of a strong radio core and two 
jets or lobes, typical of the FR\,II morphology (Fig.~\ref{images}). The overall
linear size of the source is $\sim$80\,kpc. 
The VLBA 8.4 and 15.4\,GHz observations resolve the radio core into one-sided 
core-jet structure \citep{hough2002}. The {\it Chandra}
observations of 3C270.1 revealed compact X-ray emission of the nucleus and
extended X-ray emission associated with the southern radio lobe of 3C270.1.
The extended X-ray emission probably originates in the double hotspot within
that lobe and is consistent with the inverse-Compton process
\citep{wilkes}.    

\section{Estimates of the physical parameters}
\label{sec_estim}

To estimate the basic physical parameters of the jet emission, we adopted the
approach proposed already for blazars (e.g., \citealt{bednarek1997}, \citealt{tavecchio1998},
\citealt{Katarzynski2012}). In this approach the estimations are based on specific
features observed in the spectra. For example, the position of a maximum in 
the synchrotron emission is frequently used as one of the spectral features. 
The quality of the estimations depends on the available observations and 
is usually problematic. The input parameters derived from the observations have 
relatively large uncertainties. This is somehow compensated
for by different methods used for the estimations. The procedure is 
reasonable if all the methods lead to similar values of the physical 
parameters. Finally, the estimations also depend on specific assumptions 
in the emission model.

We assume that some fraction of the emission of the 
innermost regions of BAL quasars is produced by a relatively 
small (a few pc or less) region of a jet. Therefore, to estimate the physical 
parameters of such a region, we adopt a classical scenario 
where relativistic electrons produce synchrotron and inverse-Compton 
radiation. The seed photons for the inverse-Compton scattering come from the 
synchrotron emission, as it does in the classical synchrotron self-Compton 
(SSC) process. For the sake of simplicity we assume that the emitting 
region is spherical ($R$ - radius) and homogeneous. The region is 
filled uniformly by the electrons and tangled magnetic filed 
($B$ - intensity). The particle energy distribution is described by a simple
power-law function $N(\gamma) = K \gamma^{-n}$, where $1 \leq 
\gamma \leq \gamma_{\rm max}$ and the energy is $E =\gamma m_{\rm e}c^2$.
The function $N$ gives the number of particles per unit volume at a given
energy. In principle, the source may travel with a relativistic 
velocity that implies the Doppler factor $\delta > 1$. However,
in all our estimates, we assume that $\delta=1$, because the superluminal 
motions of jet's knots were never observed in BAL quasars.

The model is described by only five free parameters ($R, B, K, n, 
\gamma_{\rm max}$). Using the well known relation $\alpha=(n-1)/2$, 
we may estimate the slope of the particle energy distribution ($n$) 
from the spectral index ($\alpha$) of the synchrotron radiation 
($F_s \propto \nu^{-\alpha}$). The SSC emission presented in 
$\nu F(\nu)$ plots usually shows two characteristic peaks. The first is
due to the synchrotron emission and the second created by the 
inverse-Compton scattering. Using the position of the synchrotron 
peak ($\nu_{s,p}$) and the level of the emission ($\nu F_s(\nu_{s,p})$) 
at this peak, we may derive a simple relation between the magnetic 
field ($B$) and the particle density ($K$) inside 
the source. An independent but similar relation can be obtained 
from the position ($\nu_{c,p}$) and the emission level 
($\nu F_c(\nu_{c,p})$) of the IC peak. Two additional relations between 
$B$ and $K$ were derived from the estimated value of the synchrotron 
self-absorption frequency ($\nu_{s,a}$) and from the assumption
about the equipartition between the magnetic field energy density 
and the particle energy density. Finally, the fifth relation that  only
constrains the magnetic field value was derived from 
the relative position of the peaks ($\nu_{s,p}$ vs. $\nu_{c,p}$). 
Detailed explanations of these estimations and derived 
formulae are given in Appendix~\ref{app_a}.

\subsection{PG 1004+130}

\begin{figure}
\includegraphics[width=8cm]{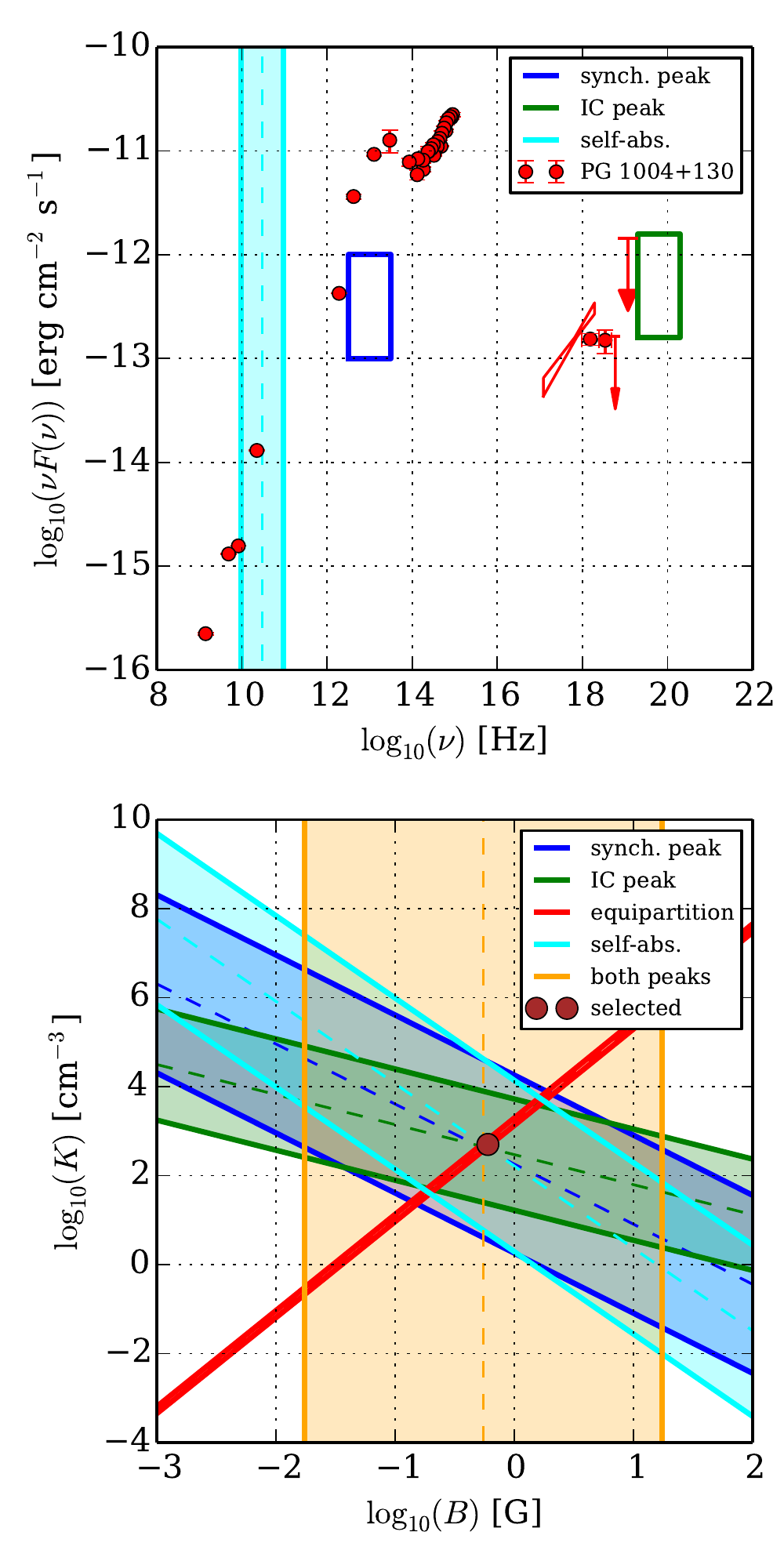}
\caption{Upper panel: All the observations collected to
estimate physical parameters of PG 1004+130 and the selected
positions of the self-absorption frequency, the synchrotron peak, 
and the IC peak with the assumed uncertainties. In the lower panel
we show the result of the estimations and the specific values 
$B = 0.7$ G and $K= 2 \times 10^{2}$ cm$^{-3}$  (a dot) selected 
for the precise calculations demonstrated in Fig. \ref{fig_1004_est}.
}
\label{fig_1004_est}
\end{figure}

To estimate the physical parameters of 1004+130, we collected
observations from radio frequencies up to the X-ray range. In the radio
domain, we focused on the emission that comes directly from the core,
excluding emission from the extended structures that are visible in the radio
maps. We assumed that the spectral index of the optically thin radio
emission is $\alpha=0.35,$ and the break due to synchrotron 
self-absorption appears at $\nu_{s,b} = 3 \times 10^{10}$ Hz. The
position of this break is not well established. Therefore, 
we assumed an uncertainty of one order of magnitude 
around the selected value. This is indicated by a vertical area
presented in the upper panel of Fig. \ref{fig_1004_est}. We consider 
that the IR and optical ranges are strongly dominated by thermal emission 
of dust and gas. Thus the position and the emission level of the
synchrotron peak are not determined precisely. We assume 
$\nu_{s,p} = 10^{13}$ Hz and $\nu F_s(\nu_{s,p})$ = $3.16 \times 10^{-13}$
erg cm$^{-2}$ s$^{-1}$ with an uncertainty of one order of magnitude
for both values. We also have to estimate the position of the 
IC peak. The X-ray observations indicate that this peak
must be located well above the {\it Chandra} energy range.
However, it is not possible to determine this peak precisely.
We assume $\nu_{c,p} = 6.3 \times 10^{19}$ Hz and 
$\nu F_c(\nu_{c,p})$ = $5 \times 10^{-13}$ erg cm$^{-2}$ s$^{-1}$
with a big uncertainty of one order of magnitude as in
the previous cases. Finally, to calculate the relations between
$B$ and $K$, we have to determine the radius. The upper limit for
this parameter can be obtained from the radio maps or from
the observed variability time scales. However, we have no 
information about the variability of PG 1004+130, and the
core observed on the radio maps may be significantly bigger
than the region where most of the emission originates.
Since there is no simple way to determine the radius, we 
selected the value $R = 2.2 \times 10^{17}$ cm that gives
a good agreement between the five independent methods 
and allows to select one particular value
of $B = 0.6$~G and $K = 5 \times 10^{2}$ cm$^{-3}$ for 
more precise simulations that we discuss in the next section. 
Also, in this case we assume one order of magnitude uncertainty 
in the $R$ value. The selected values with the uncertainties 
and the resulting estimations are presented in  Fig. 
\ref{fig_1004_est}. The assumed uncertainties 
of the input parameters give an area instead of a simple 
functional relationship derived in Appendix~\ref{app_a}.

\subsection{3C 270.1}

\begin{figure}
\includegraphics[width=8cm]{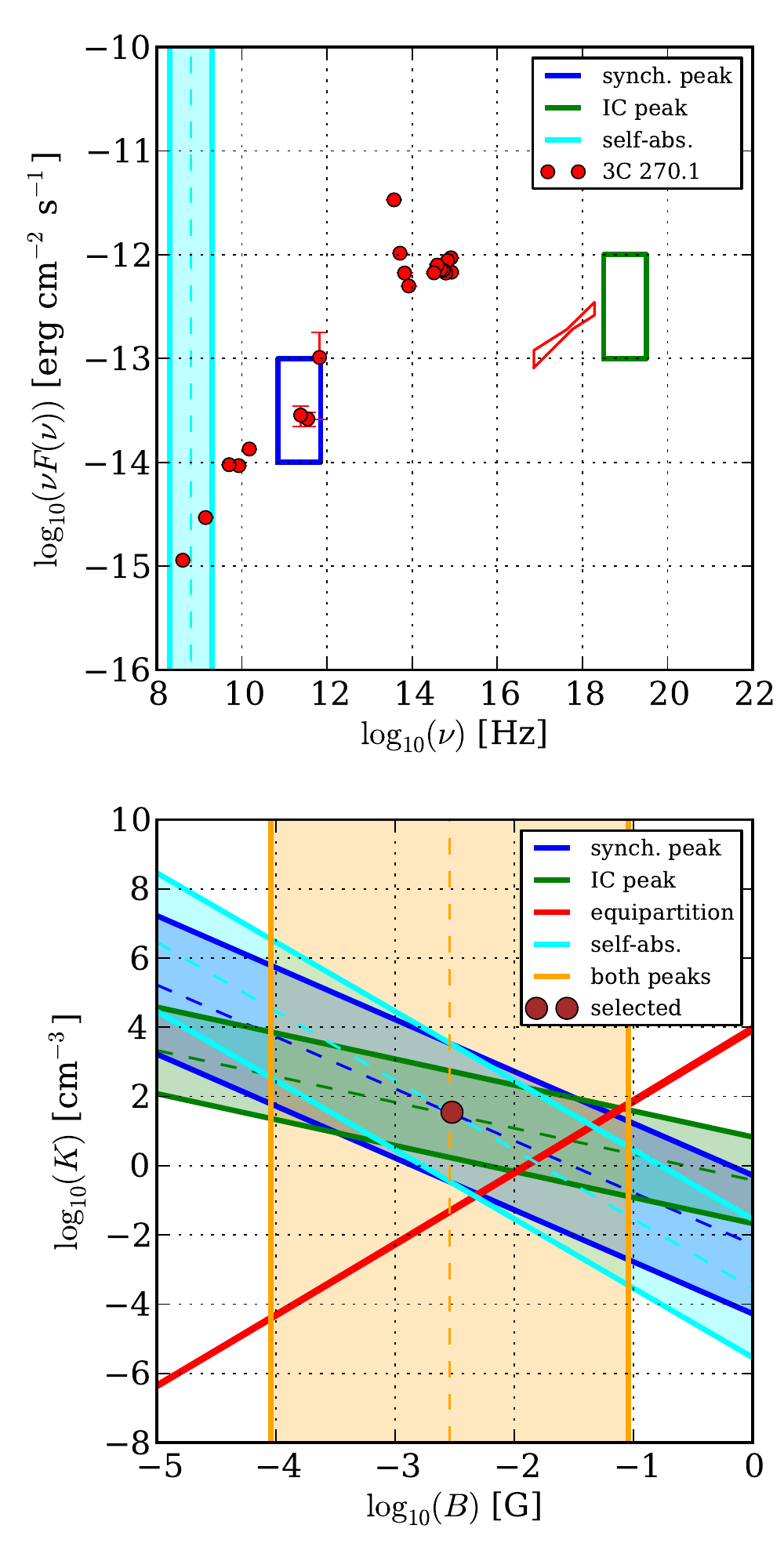}
\caption{Observations of 3C 270.1 collected to estimate 
physical parameters and the selected input parameters 
with the assumed uncertainties (upper panel). The
lower panel shows the result of the estimations and 
the selected values of $B = 0.003$ G and $K= 35$ cm$^{-3}$.
There is no agreement between the estimation 
based on the equipartition and the other methods.
}
\label{fig_3C270_est}
\end{figure}

To estimate the physical parameters in 3C 270.1, we conducted
 similar procedure to the one for PG 1004+130. We chose the spectral
index of $\alpha = 0.5$, the self-absorption frequency $\nu_{s,b} 
= 6.3 \times 10^{8}$ Hz, position of the synchrotron peak $\nu_{s,p}
 = 2.24 \times 10^{11}$ Hz, $\nu F_s(\nu_{s,p})$ = $3.16 \times 10^{-14}$
erg cm$^{-2}$ s$^{-1}$, position of the IC peak $\nu_{c,p} = 10^{19}$ Hz, 
$\nu F_c(\nu_{c,p})$ = $3.16 \times 10^{-13}$ erg cm$^{-2}$ s$^{-1}$,
and the radius $R = 5 \times 10^{19}$ cm. For all above parameters,
we assumed an uncertainty of one order of magnitude around the
chosen values. The observations used for the estimations, 
the values of the input parameters with assumed uncertainties, 
and the results of the estimations are presented in Fig.~\ref{fig_3C270_est}. 
The main difference between this estimation 
and the calculations performed for PG 1004+130 is that there 
is no agreement between all the used methods. 

The estimation that 
assumes equipartition between the magnetic field energy density 
(Eq. \ref{eq_mag_den}) and the particle energy density 
(Eq. \ref{eq_par_den}) differs significantly from other 
estimations. It seems that there is no such equipartition in 
this particular source ($U_{e} \gg U_{B}$). This is also indicated
by the fact that the IC peak is about one order of magnitude
higher than the synchrotron peak. The equipartition would 
require $\nu F_{\nu}$ peaks at about the same level. 
In such a case it would be possible to obtain the agreement between the five 
estimation methods. However, this is not obtained. The lack of the 
equipartition may be caused by the efficient particle acceleration 
inside the emitting region. The acceleration by the first-order
Fermi process at a shock wave inside the jet is frequently 
postulated as a source of energetic particles.

\section{Modelling}

We assume that most of the radio and the X-ray emission is
produced by a relatively small region of a jet. Therefore,
we adopt a very simple model to explain this emission. 
On the basis of this model, we derived five independent 
methods of  estimating the magnetic field and the particle
density inside the emitting zone. Using the estimated 
values and the more precise computation scheme proposed 
by \citealt{Katarzynski01}, we show how to explain the
radio and X-ray emission. To explain the emission in
 the IR, optical, and UV ranges, we also have to postulate 
other emission processes.

In the IR range the emission is probably dominated by the thermal radiation
of the dust inside the host galaxy \citep{hughes1993, haas1998}. In principle, this kind of 
emission can be modelled by the grey-body spectrum. However, it
appears necessary to assume different temperatures and emission
levels for the emitting dust. Therefore, we describe this emission by
\begin{equation}
\nu L(\nu) = \nu_p L(\nu_p) f(\nu)/f(\nu_p),
\end{equation}
where
\begin{equation}
f(\nu) = \int_{T_{\rm min}}^{T_{\rm max}} T^s B_{gb}(T, \nu) d T.
\end{equation}
Here, $B_{gb}$ is the grey-body spectrum (e.g. \citealt{Kunert09}) and
$T_{\rm min}$ and $T_{\rm max}$ are the minimum and maximum
temperatures of the dust. The spectrum is described also by
the luminosity $L(\nu_p)$ at the peak ($\nu_p$) and
the slope $s$ that defines the difference in the intensity of the
emission. All these parameters can be derived directly from the
observations.

Finally, we assume that the optical and UV emission comes mostly from the accretion disk, although the 
contribution from stars in the host galaxy cannot be excluded. Some fraction of the X-ray emission may
come from the corona of the accretion disk. To simulate such radiation, 
we used the model proposed by \citealt{Janiuk00}.
In our computations, the fraction of energy dissipated 
in the corona, $f_{c}$, is not a free parameter but is computed   
self-consistently by solving a closed set of equations, and it has a value
between $\sim 0.1$ and 1.0, with the profile that is increasing with radius.
Also, all the measurable quantities, i.e. the flux ratio, the spectral
slopes, and the extension of the spectrum into the gamma-ray band, result
from the model, including the trends for the change in these
quantities with the accretion rate. The spectral shape is affected by both  
viscosity $\alpha_{\rm d}$ and accretion rate $\dot m$. In general, for a rather 
low viscosity, large $\dot m$ have an effect on the softening of the
spectrum. 
For large $\alpha_{\rm d}$, this dependence is more complex, because the electron
temperature decreases with $\alpha_{\rm d}$.

The non-rotating black hole model is taken here for simplicity, although 
in the case of the radio-loud QSOs, the rotating black hole is usually 
considered. The spin of the black hole would result in a shifting of the
inner radius of the disk-corona system closer to the black hole. 
However, the coronal emission is dominated by the outermost parts of the 
corona, and the X-ray spectrum would not change because of the black hole
rotation.
Therefore the outcome of the modelling is practically unaffected by the value
of the black hole spin.

\subsection{Spectrum of PG 1004+130}

\begin{figure}
\includegraphics[width=8cm]{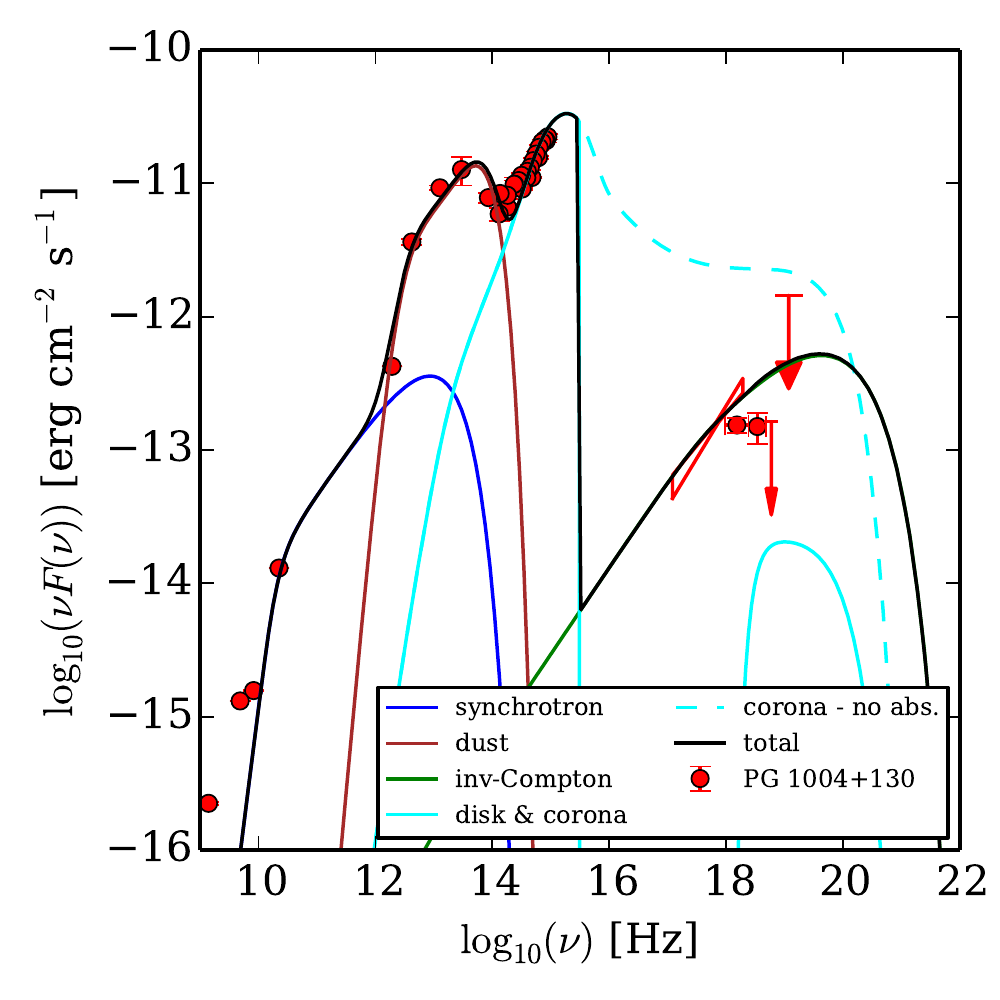}
\caption{Spectrum of PG 1004+130, from radio frequencies
up to X-rays. To explain this spectrum it is necessary
to simulate four processes: the synchrotron radiation
of the jet (radio $\to$ IR), grey-body emission of dust 
inside the host galaxy (IR $\to$ optical), thermal emission 
of the accretion disk (optical $\to$ UV), and the inverse-Compton 
scattering inside the jet (X--rays). 
In addition, we add a weak hard X-ray emitting corona, which is absorbed 
by circumnuclear gas with a large column density on the line of sight.
}
\label{fig_1004_mod}
\end{figure}

The radio spectrum of the core of PG 1004+130 is inverted 
$\alpha \simeq -0.5$ ($F \propto \nu^{-\alpha}$). This indicates 
that the entire radio emission is significantly self-absorbed.
This requires a relatively compact source. Indeed, our estimations
show that the source radius should be about 0.1 pc. What is
interesting is the equipartition between $U_B$ and $U_{e}$
in this compact object. In a more extended source, it would be impossible 
to obtain equipartition, because it would require many more 
particles to compensate for the low efficiency of the IC 
scattering, and therefore $U_{e}$ would be much greater 
than $U_B$.

The emission from IR up to UV range can be explained by the
emission of dust and the accretion disk. The IC scattering 
inside the jet explains the {\it Chandra} observations well
(Fig. \ref{fig_1004_mod}). Note, that there is no agreement
between the {\it Chandra} and {\it NuSTAR} observations. The
spectrum slope and the level of the emission are different.
Therefore, we focus on the {\it Chandra} observations that
seem to be more precise. 

However, the theoretically obtained spectrum produced by the accretion disk and corona
around a highly massive black hole (dotted blue line in Fig. \ref{fig_1004_mod}), greatly exceeds the level of UV and X-ray emission observed in the data for PG\,1004+130. Thus, in the modelling
of the disk and corona emission we consider the internal
absorption of X-ray photons produced in the corona and postulated by \citet{luo13}, based on their 
{\it NuSTAR} observations.
As a parameter, we assumed the hydrogen column density of $7 \times 10^{24}$ 
cm$^{-2}$, and we fitted the effective cross-section 
as a function of energy in the interstellar  medium, 
adapted from \citet{cruddace1974}. In addition,
we take the electron scattering of hard X-ray photons
in the absorbing cloud into account. After these corrections, where the corona emission is absorbed, we get a satisfactory fit (Fig. \ref{fig_1004_mod}).

\subsection{Spectrum of 3C 270.1}

\begin{figure}
\includegraphics[width=8cm]{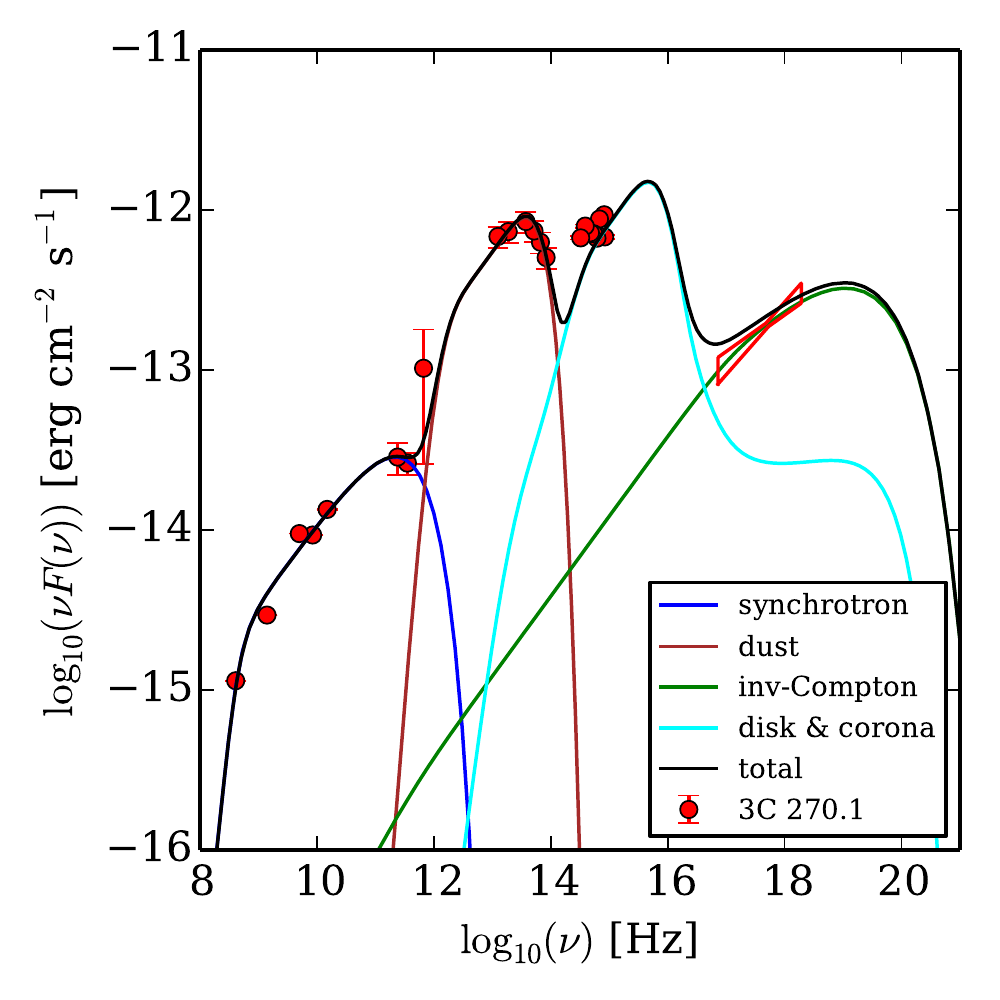}
\caption{Emission of 3C 270.1 in a wide frequency
range calculated as the synchrotron emission of the jet (radio $\to$ IR), 
thermal emission of stars in the host galaxy (IR $\to$ optical), 
the radiation of the accretion disc and corona (UV $\to$ X-ray), 
and the IC emission generated by the jet (X--rays).
}
\label{fig_3C270_mod}
\end{figure}

This source appears to be significantly different from PG~1004+130.
Most of the radio emission is produced by the optically thin source. The
spectral index has a classical value of $\alpha=0.5$,
and the self absorption seems to be significant only at the lowest 
observed frequencies. Therefore, the source is relatively
large, a few tens of parsec. The IR and optical emission can be
explained by the radiation of the dust in the galaxy. The UV
emission can only be produced by the accretion disk. Since the
X-ray emission is relatively high, the contribution from the 
corona is almost negligible. The strong X-ray flux indicates that
equipartition is not possible. The parameters 
used to describe the spectrum of 3C 270.1 are given in Table \ref{tab_params},
and the calculated spectra are presented in Fig. \ref{fig_3C270_mod}.

\begin{table}
\caption{Best fit parameters.}
\begin{tabular}{lllll}
\hline
params          & PG 1004+130           & 3C 270.1            & unit       & source\\
\hline
$R$                 & 2.2 $\times 10^{17}$  & $5 \times 10^{19}$  & cm         & jet comp.\\
$B$                 & 0.6                   & 0.003               & G          & jet comp.\\
$K$                 & 5.0 $\times 10^{2}$     & 35                  & cm$^{-3}$  & jet comp.\\
$\alpha$            & 0.35                  & 0.5                 &            & jet comp.\\
$\gamma_{\rm max}$  & $3.2 \times 10^{3}$   & $1.2 \times 10^{4}$ &            & jet comp.\\
$\nu_p L(\nu_p)$    & $2.3 \times 10^{45}$  & $1.3\times 10^{46}$ & erg/s      & dust \\
$T_{\rm min}$       & 50                    & 45                  & K          & dust \\
$T_{\rm max}$       & $1.5\times 10^{3}$                   & $2.3\times 10^{3}$  & K          & dust \\
$s$                 & -4.4                  & -4.5                &            & dust \\
$M_{BH}$            & $1.2 \times 10^{9}$   & $1.7 \times 10^{9}$ &$M_\odot$   & disk \& cor. \\
$\dot{m}$           & 0.08               & 0.2                 & $M_\odot/{\rm yr}$            & disk \& cor. \\
$\alpha_{\rm d}$    &0.03                  & 0.1                 &            & disk \& cor. \\
\hline
\end{tabular}
\label{tab_params}
\end{table}

\section{Discussion}

The main aim of this paper is to explain the nuclear spectra
of the selected broad absorption line (BAL) quasars in the whole observed energy range. 
In the literature, the origin of the X-ray emission in BAL quasars is still a widely discussed issue.
Given the multiwavelength data and the interferometric radio properties of both PG\,1004+130 and 3C\,270.1, which are BAL quasars, 
it appears that the non-thermal, inverse-Compton
emission from the innermost parts of the radio jet can account for a significant part of the observed X-ray emission in these objects. In the case of 3C\,270.1, the jet X-ray emission is very strong,
without giving the possibility of any contribution from the corona. A larger thermal corona contribution is possible in the model of less
radio luminous PG\,1004+130. However, in reality PG\,1004+130 is much less X-ray luminous than 3C\,270.1, which can be caused probably by the absorption of Compton-thick shielding gas \citep{luo13}.
How are these results related to the general discussion of the X-ray properties of BAL quasars?

\subsection{X-ray properties of BAL quasars}

In the accretion disk/wind model proposed by \citet{Murray1995} and then explored by \citet{Proga2000}, the 
outflows are launched from the accretion disk at the distance $<10^{17}$ from the supermassive black hole and 
radiatively driven by the UV line pressure. The X-ray emission from the innermost parts of the accretion disk and corona are
absorbed interior to the UV BALs probably by the shielding gas. Its origin and geometry are still uncertain, but 
it is thought to be the reason for the X-ray weakness of BAL quasars. It has been suggested that the disk outflows are 
present in all quasars, but the effect of their presence, namely the BALs and X-ray absorption, can only
be detected when looking at specific, mid-inclination angles \citep{elvis00}. 

Generally, after accounting for the intrinsic absorption, the X-ray properties of BAL quasars are consistent with those of 
unabsorbed quasars \citep{gallagher2006, stalin2011}. Nevertheless, $17\%-40\%$ of BAL quasars still remain X-ray weak \citep{luo13}. Recently, \citet{leah14} have
shown that when correcting for the magnification of X-ray emission via gravitational lensing by the central black hole, viewed at large inclination
angles, the BAL quasars appear even more intrinsically X-ray weak. The X-ray absorption in BAL quasars is complex, often
requiring fitting of an ionized or partially covering absorber, or a mixture of both. Another pure constrained issue is the relationship between
UV and X-ray absorbing material. Some kind of connection is suggested by the correlation between X-ray weakness of BAL quasars and the absorption
strength and maximum velocity of the UV C\,IV BALs \citep{strebl10}.
The spectroscopic and variability studies trying to characterize the UV absorber in BAL quasars generally 
indicate the complex behaviour of wind outflows. BALs in radio-quiet and radio-loud quasars show depth changes in a wide range from minor to 
complete disappearance \citep{filiz2012}. This can be caused, for example, by changing the location of an absorber along the line of sight \citep{capellupo2013}. 
Recent observations of BAL variability in the sample of radio-loud BAL quasars support the orientation dependence of the observed outflow \citep{welling2014}.

The radio studies of BAL quasars indicate that strong UV C\,IV BALs are associated with lower values of the radio-loudness parameter \citep{welling2014, kun14}.
Since this parameter is thought to be a good indicator of orientation \citep{kimball2011}, its low values may imply large viewing angles and consequently
the orientation dependence of BAL quasars. However, there is no correlation between the radio-loudness parameter and the BI, since a large
span of values occurs in each bin of the radio-loudness parameter. Therefore, as pointed out by \citet{kun14}, the orientation is only one of the factors 
influencing absorption. 

Another interesting feature of radio-loud BAL quasars is that they are less X-ray weak than radio-quiet BAL quasars compared to non-BAL objects \citep{miller2009}. This is probably connected with the radio emission. The simple scenario suggests that the X-ray continuum in radio-loud BAL quasars
consists of both accretion disk-corona and jet-linked X-ray emission coming from the innermost parts of the radio jet \citep{wang2008,Kunert09, miller, miller2009,berrington}. 
However, the distinction of the particular components (inner and outer parts of the jet, radio core) of radio-loud BAL quasars is often impossible, because they
remain unresolved even in the high-resolution observations \citep{jiang2003, kunert2007, kunert2010, liu2008, monte2008, gawron2011, bruni2012, 
bruni2013, hayashi2013}. 
The large scale BAL quasars, PG\,1004+130, and 3C\,270.1 discussed in this paper have clearly resolved radio structures with radio and X-ray core detection. The majority of the 
X-ray emission is concentrated precisely in the centers of these objects.

\begin{figure}
\centering
\includegraphics[width=8cm, height=6cm]{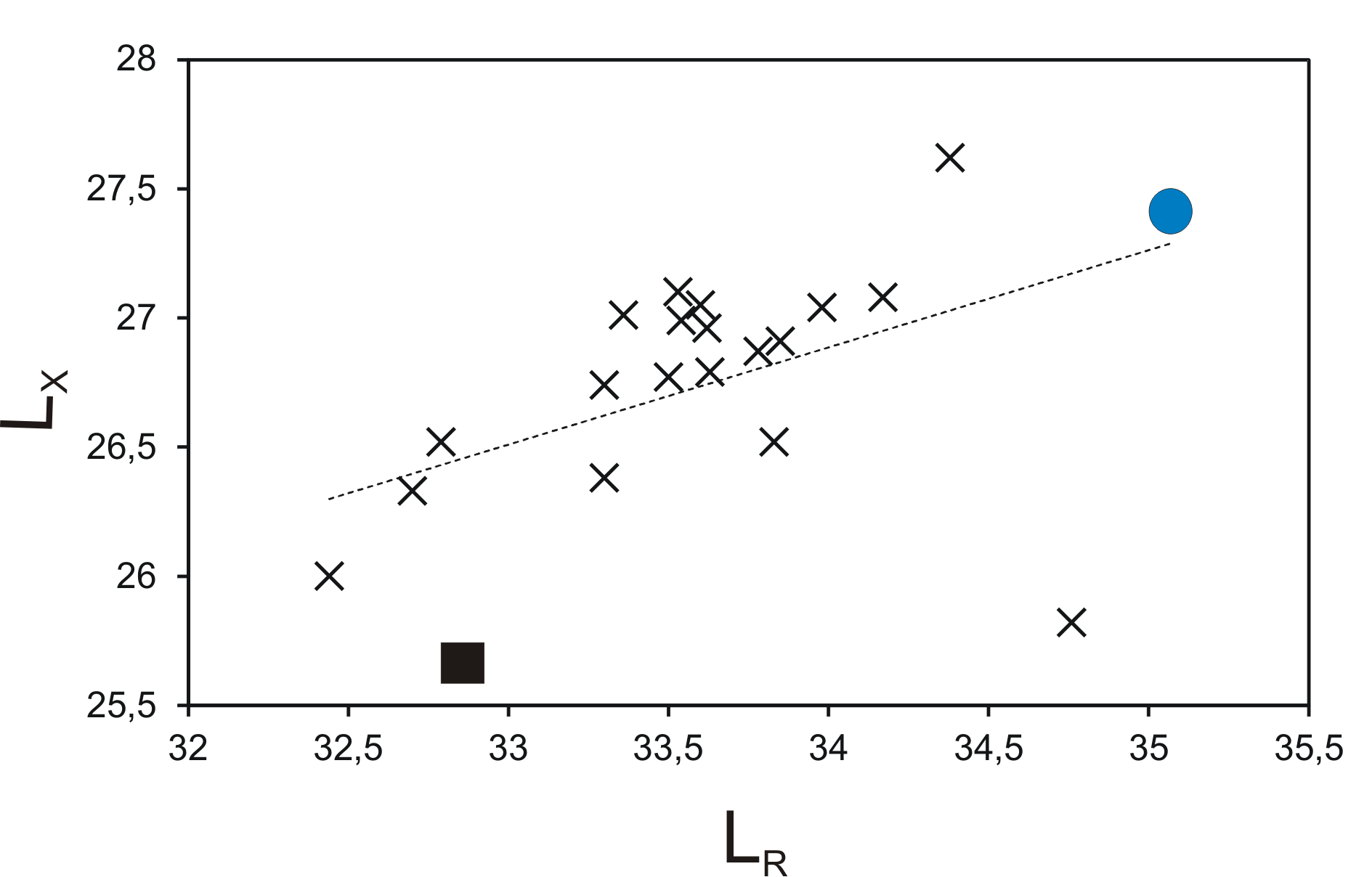}
\caption{X-ray - radio luminosity correlation for radio-loud BAL quasars. Data are taken from \citet{miller2009}.
Luminosities are in the units of log $\rm ergs\,s^{-1}\,Hz^{-1}$ at rest frame frequencies of 5\,GHz and 2\,keV. BAL
quasars studied in this paper are indicated with a black square (PG\,1004+130) and with 
a blue circle (3C\,270.1).}
\label{correlation}
\end{figure}

\subsection{Two ends of the radio vs. X-ray luminosity dependence}

Both PG\,1004+130 and 3C\,270.1 belong to the population of broad absorption line (BAL) quasars but differ in many aspects. 
Their radio and X-ray luminosities place them on two different ends of the X-ray - radio dependence for radio-loud BAL quasars (Fig.~\ref{correlation}).
{\it XMM-Newton} observation of PG\,1004+130 shows only minimal intrinsic absorption, while later {\it Chandra} X-ray observations of this source
revealed modest X-ray absorption ($\rm N_{H}\sim 10^{22}\,cm^{-2}$) with partial-covering absorbed model \citep{miller}. 
Recently, PG\,1004+130 has been also detected in the softer {\it NuSTAR} band (4-20\,keV) but not in its harder bands ($>20\,keV$), which may imply significant Compton-thick obscuration ($\rm N_{H}\sim 7\times10^{24}\,cm^{-2}$). An alternative explanation of the X-ray weakness of PG\,1004+130 invokes a specific mechanism that weaken the X-ray emission of the disk and corona. If the first suggestion is true, the lack of the
strong Fe\,K$\alpha$ line emission is surprising and argues against the scenario in which the nucleus of PG\,1004+130 is heavily absorbed. However, the Fe\,K$\alpha$ line emission could be diluted by the jet-linked X-ray emission \citep{luo13}. Such a phenomenon has already been observed in the radio-loud AGNs \citep{grandi2006}.
The brighter of our two quasars, 3C\,270.1, shows no evidence of intrinsic absorption, but the  Fe\,K$\alpha$ line is marginally detected \citep{wilkes}.

There is also a difference between PG\,1004+130 and 3C\,270.1 in the value of the BI, which quantify the UV C\,IV absorption. Each BAL quasar has to have $\rm BI\,[km\,s^{-1}]>0$, which means $\rm C_{IV}$ absorption troughs that are at least $\rm 2000\,km\,s^{-1}$ wide and at least 10\% below the continuum at
maximum depth \citep{Weymann1991}.
The BI value amounts to 850 and 52.5 for PG\,1004+130 and 3C\,270.1, respectively \citep{wills99, gibson2009}. As already mentioned, 
the strong UV absorption seems to be associated with lower values of radio-loudness parameter, log\,$\rm R_{I}<1.5$ and thus large viewing angles \citep{welling2014, kun14}. 
The values of the radio-loudness parameter, log\,$\rm R_{I}$, defined as the radio\,(1.4\,GHz)-to-optical\,(i-band) 
ratio of the quasar core \citep{kimball2011} amounts to 0.7 and 2.8 for PG\,1004+130 and 3C\,270.1, respectively. We can thus suspect stronger UV absorption in PG\,1004+130 than in 3C\,270.1.

The difference in radio properties of both quasars is also significant. 3C\,270.1 is a core-dominated FR\,II object on VLA scales. 
The high-resolution 8.4 and 15.4\,GHz VLBA observations resolve the core of 3C\,270.1 into a one-sided core-jet structure. \citet{hough2002} report
the presence of a strong (26\,mJy at 15.4\,GHz) and a small ($\rm \sim 0.2\,mas$) radio core. There is no information about the parsec scale structure of PG\,1004+130, but 
the VLA large scale morphology is not typical. PG\,1004+130 has been described as a hybrid object with two different FR structures on both sides of the source \citep{gopwii}.
Recent studies of hybrid objects \citep{gawron2006, cegla2013} indicate that the interstellar medium can be responsible for their complex structures, but 
the properties of their central engines can also be an important factor here. Lower luminosity AGNs may develop into diffuse, large-scale structures 
because their weak jets are disrupted before escaping their host galaxies.

According to the analysis performed by \citep{willott1999}, the radio luminosity of a large-scale radio source is approximately proportional to the jet power. Adopting the 
modifications made by \citep{sikora2013}, we use their equation $\rm P_{jet}\sim 10^{2}\,\nu_{1.4\,GHz}\,L_{1.4\,GHz}\,(f/3)^{3/2}\,W\,Hz^{-1}$, where factor 
{\it f} is in the range 1- 20 (we adopted f=10), to calculate jet power of the selected BAL quasars, and these amount to $\rm 10^{38}\,W\,Hz^{-1}$ and $\rm 10^{40}\,W\,Hz^{-1}$ for PG\,1004+130 and 3C\,270.1, respectively.
Our modelling of the spectral energy distribution of both quasars indicates that the contribution of the jet-linked X-ray emission is possible in both  the strong
and weak radio sources, but its fraction seems to scale with jet or radio power. Since the majority of BAL quasars are weak compact radio sources without prominent jets 
\citep{bruni2012,bruni2013,kun14},
such scaling could be responsible for the difference in X-ray emission between radio-loud BAL quasars and non-BAL objects. Additionally \citet{miller2011} suggest that beaming
can increase the dominance of jet-linked X-ray emission at low inclinations in radio-loud BAL quasars.

\section{Summary}
PG\,1004+130 and 3C\,270.1 studied in this paper belong to the same class of broad absorption line (BAL) quasars but differ in radio, UV, and X-ray properties.
We call them two ends of the radio versus X-ray luminosity correlation. They are large enough to be resolved by the high-resolution radio observations, which is rare
among the BAL quasars. The clearly detected radio cores with multi-wavelength data allowed us to model their nuclear spectral energy distributions assuming 
that it contains both disk/corona-linked and jet-linked components. We argue that in
the case of radio luminous 3C\,270.1 a non-thermal, inverse-Compton emission from the innermost parts of the radio jet can account for a large
amount of the observed X-ray emission. The disk/corona-linked X-ray emission is also present but is dominated by the jet-linked X-ray emission.
In the case of lobe-dominated PG\,1004+130, the radio core is probably too weak to produce any significant
part of observed X-ray emission. A large contribution from the X-ray emission of the accretion disk and corona is possible in our model. However, in reality the observed X-ray emission of PG\,1004+130 is much weaker, and such large suppression can be caused by the absorbing gas present near the black hole. As a consequence, a large part of the X-ray emission observed in PG\,1004+130 must arise in the radio jet.

The results of our modelling show that the jet-linked X-ray emission is present in both the strong and weak radio sources, but its fraction scales with jet or radio power.
Orientation could be an additional factor influencing the observed emission. It is more difficult to constrain the fraction of the disk/corona-linked X-ray emission. 
It could be orientation dependent, but recent results suggest that it is a simplification and the nature, and behaviour of X-ray absorber is more complex.

\section*{Acknowledgements}
We thank Bo\.zena Czerny for discussion and constructive comments.

This work was supported by the National Science Centre under grant DEC-2011/01/D/ST9/00378, and partially by 
grant DEC-2012/05/E/ST9/03914.

The National Radio Astronomy Observatory is a facility of the National
Science Foundation operated under cooperative agreement by Associated 
Universities, Inc. The VLA data was calibrated using NRAO's "VLA data 
calibration pipeline" in AIPS.

This research has made use of the NASA/IPAC Extragalactic Database (NED), which is operated by the Jet Propulsion Laboratory, California Institute of Technology, under contract with the National Aeronautics and Space Administration.

Funding for the SDSS and SDSS-II has been provided by the Alfred P. Sloan Foundation, the Participating Institutions, the National Science Foundation, the U.S. Department of Energy, the National Aeronautics and Space Administration, the Japanese Monbukagakusho, the Max Planck Society, and the Higher Education Funding Council for England. The SDSS Web Site is http://www.sdss.org/.

The SDSS is managed by the Astrophysical Research Consortium for the Participating Institutions. The Participating Institutions are the American Museum of Natural History, Astrophysical Institute Potsdam, University of Basel, University of Cambridge, Case Western Reserve University, University of Chicago, Drexel University, Fermilab, the Institute for Advanced Study, the Japan Participation Group, Johns Hopkins University, the Joint Institute for Nuclear Astrophysics, the Kavli Institute for Particle Astrophysics and Cosmology, the Korean Scientist Group, the Chinese Academy of Sciences (LAMOST), Los Alamos National Laboratory, the Max-Planck-Institute for Astronomy (MPIA), the Max-Planck-Institute for Astrophysics (MPA), New Mexico State University, Ohio State University, University of Pittsburgh, University of Portsmouth, Princeton University, the United States Naval Observatory, and the University of Washington.

\appendix

\section{}
\label{app_a}

We derived four independent methods of estimating the relation
between magnetic field ($B$) and particle density ($K$) for 
the simple SSC model used in this work. In addition we provide a simple method of constraining the magnetic field value.

\subsection{Synchrotron peak}

According to our assumption, the particle energy distribution is
described by a power-law function
\begin{equation}
\label{eq_part_energy_spec}
N(\gamma) = K \gamma^{-n} \;\; {\rm for} \;\; 1 \leq \gamma \leq \gamma_{\rm max} \;\; [{\rm cm}^{-3}].
\end{equation}
For such a simple distribution, the synchrotron emissivity in the
source's comoving frame can also be approximated by 
a power-law relation
\begin{equation}
j_s(\nu') = \frac{C_s(n)K B^{\alpha+1}}{4 \pi}(\nu')^{-\alpha} 
\;\; \left[\frac{\rm erg} {{\rm s} \; {\rm cm}^{3} \; {\rm sterad} \; {\rm Hz}}\right],
\end{equation}
where $\alpha = (n-1)/2$ and the coefficient
\begin{eqnarray}
C_s(n) & = & \frac{4 \sqrt{3} \pi e^3}{m_e c^2} \left( \frac{3e}{2\pi m_e c} \right)^{\frac{n-1}{2}} \Gamma\left(\frac{3n-1}{12}\right) \Gamma\left(\frac{3n+19}{12}\right)\nonumber\\
     & \times & \Gamma \left( \frac{n+5}{4}\right)  /  \left( 8 \sqrt{\pi} (n+1) \Gamma((n+7)/4) \right)
\end{eqnarray}
(e.g. \citealt{Ginzburg65}). The coefficient above is quite complex.
However, it can be approximated very well by a simple polynomial formula
\begin{equation}
\log_{10}(C_s) \simeq 0.08n^2 + 2.76n - 24.27
\end{equation}
that is precise enough ($\Delta C_s < 2\%$ for $1.5 \leq n \leq 4$) for our
estimations. All the formulas we present in this appendix are universal, 
applicable for any synchrotron and/or SSC source. Therefore, our goal was
to provide as simple equations as possible. In all the cases we provide 
a simple polynomial approximation of more complex coefficients.

The intensity of the surface emission of spherical, optically thin 
source is given by 
\begin{equation}
\label{eq_synch_int}
I_s(\nu') = \frac{4}{3} j_s(\nu') R \;\; [{\rm erg} \; {\rm s}^{-1} \; {\rm cm}^{-2} {\rm sterad}^{-1} {\rm Hz}^{-1}],
\end{equation}
and the flux is defined by
\begin{equation}
F_s(\nu) = \pi \frac{R^2}{D_L^2} (1+z) \delta^3 I_s(\nu') \;\; \left[{\rm erg}/{\rm cm}^2\right],
\end{equation} 
where: $\delta$ is the Doppler factor, z the redshift, $D_L$ the luminosity distance. 
The frequency transformation is given by $\nu = \nu' \delta / (1+z)$.
Using these formulae we may write our first $B \to K$ relation
\begin{equation}
K(B) = D_s(n) \frac{F_{s,p}}{f_s},
\end{equation}
where
\begin{equation}
f_s = \frac{C_s}{3} \frac{R^3}{D^2_L} (1+z)^{1-\alpha} \delta^{3+\alpha} B^{\alpha+1}\nu^{-\alpha}_{s,p}
,\end{equation}
and 
\begin{equation}
D_s(n) = 0.61 n^4 - 5.97 n^3 + 22.2 n^2 -38.4 n + 27.6
\end{equation}
is the correction coefficient. We use a simple power-law
approximation of the synchrotron spectrum that differs
from precisely calculated spectrum, especially at the
$\nu F_s(\nu) $ peak, where the precise spectrum has
a curved shape. Therefore, it is necessary to introduce 
an adequate correction. We have calculated the difference
between the precise and the approximated spectrum for
different values of $n$ ($1.5 \leq  n < 3$) 
and expressed this difference by the polynomial 
presented above. We note that $F_{s,p}$ and $ \nu_{s,p}$ are
observable quantities, the Doppler factor in all our
estimations is assumed to be $\delta=1,$ and the radius
is constrained from the radio maps.

\subsection{Inverse-Compton peak}

The inverse-Compton emissivity in our simple SSC scenario
can also be approximated by a power-law function
\begin{equation}
j_c(\nu') = C_c(\alpha) K^2 B^{\alpha+1} (\nu')^{-\alpha} 
\;\; \left[\frac{\rm erg} {{\rm s} \; {\rm cm}^{3} \; {\rm sterad} \; {\rm Hz}}\right],
\end{equation}
where $\log_{10}(C_c) = 6.58 \times 10^{-4}\alpha - 46.2.$
This approximation is valid only in the Thompson 
regime of the scattering. The intensity, flux, and frequency 
transformations are identical as in the case of the synchrotron 
emission. Thus, second $B \to K$ relation is
\begin{equation}
K(B) = \sqrt{D_c(n) \frac{F_{c,p}}{f_c}},
\end{equation}
where by analogy
\begin{equation}
f_c = \frac{C_c}{3} \frac{R^3}{D^2_L} (1+z)^{1-\alpha} \delta^{3+\alpha} B^{\alpha+1}\nu^{-\alpha}_{c,p},
\end{equation}
and the correction coefficient is given by
\begin{equation}
D_c(n) = -5.1 n^3 + 37.3 n^2 - 100.6 n + 102.5.
\end{equation}
The other assumptions for this estimation are identical, as in the case of the synchrotron emission.

\subsection{Equipartition}

The other relation was derived from the assumption
about the equipartition ($U_B = U_e$) between the 
magnetic field energy density 
\begin{equation}
U_B = \frac{B^2}{8\pi}
\label{eq_mag_den} 
\end{equation}
and the particle energy density
\begin{equation}
U_e = m_{\rm e} c^2 \int_1^{\gamma_{\rm max}} \gamma N(\gamma) d \gamma,
\label{eq_par_den} 
\end{equation}
which gives
\begin{equation}
K(B) = 
\left\{
\begin{array}{ll}
\displaystyle \frac{B^2}{8 \pi m_{\rm e} c^2 \ln \gamma_{\rm max}}, & n=2\\
\displaystyle \frac{(2-n)B^2}{8 \pi m_{\rm e} c^2 \left(\gamma_{\rm max}^{2-n}-1\right)}, & n \neq 2\\
\end{array}
\right.,
\end{equation}
where the Lorentz factor that characterize the maximal energy of 
the particles was obtained from the synchrotron peak frequency
\begin{equation}
\label{eq_gamma_max}
\gamma_{\rm max} = \sqrt{\frac{\nu_{s,p} \: \delta}{(1+z) P_s(n) B}},
\end{equation}
where $P_s(n) = -1.9 \times 10^5 n^3 + 8.4 \times 10^5 n^2 
               - 1.9 \times 10^6 n +3.5 \times 10^6$
was obtained for the specific case of the synchrotron peak
created by an abrupt cutoff in the particle energy 
distribution (Eq. \ref{eq_part_energy_spec}).
               
\subsection{Self-absorption frequency}               
               
We assumed that the source is optically thin (Eq. \ref{eq_synch_int})
at the frequencies (IR to X-ray) where the peaks are observed. 
This is very reasonable assumption that simplifies the calculations.
However, at radio frequencies, the synchrotron self-absorption 
process may drastically change the slope of the spectrum. The low-frequency,
self-absorbed emission will have a constant slope $\alpha=5/2$, 
whereas the spectral index of the optically thin part will depend on
the particle energy slope ($\alpha=(n-1)/2$). This gives
a characteristic break in the spectrum at the frequency that depends 
on $B, K, R,$. The coefficient that describes the absorption for the
source with a power-law particle energy distribution can 
also be approximated by a power law formula
\begin{equation}
k_s(\nu') = B_s(n) K B^{1.5+\alpha} (\nu')^{-(\alpha + 5/2)} \;\; \left[{\rm cm}^{-1}\right]
\end{equation}
where $\log_{10}(B_s) = 0.14n^2 + 2.43n + 6$ for $1.5 \leq n \leq 4$.
Assuming that at the break frequency ($\nu_{s,b}$) the optical depth 
($\tau = R k_s$) is equal unity, we may obtain another relation
between $B$ \& $K$
\begin{equation}
K(B) = \frac{((1+z)\nu_{s,b}/\delta)^{\alpha+2.5}}{B_s(n) R B^{\alpha+1.5}}.
\end{equation}
               
\subsection{Relative peak positions}               
               
Finally, a simple constraint for only the magnetic field
can be obtained from the formula \ref{eq_gamma_max}
\begin{equation}
B = \frac{(1+z)\nu_{s,p}}{P_s(n)\gamma^2_{\rm max} \delta}
\end{equation}
where this time $\gamma^2_{\rm max} = \nu_{c,p}/(P_c(n) \nu_{s,p})$
and $P_ c(n) = -0.48n+1.3$. The maximum energy ($\gamma_{\rm max}$) 
was derived from relative position of the peaks. According to
the basic theory of the IC scattering, on average 
$\nu'_{c,p} \simeq (4/3) \gamma^2_{\rm max} \nu'_{s,p}$.
However, our simulations show that in the case of simple
power-law particle energy distribution with a sharp cut-of,f 
it is more precise to use the coefficient $P_c$ that
depends on the slope instead of the factor 4/3, 
in the above formulae.

\end{document}